\documentclass[a4paper]{article}
\pdfoutput=1
\usepackage{hyperref}
\hypersetup{
  pdfinfo={
    Title={Your Title Here},
    Author={Your Name Here},
    Subject={If you want to put something here, do so},
    Keywords={Add some keywords if you feel so inclined}
  }
}
\usepackage{pdfpages}
\usepackage{graphicx} 

\title{Supplementary Material: A quantized anomalous Hall effect above 4.2 K in stacked topological insulator/magnet bilayers }
\usepackage{fullpage}
\usepackage{amsmath}
\usepackage{xcolor}%
\usepackage{natbib}
\usepackage{placeins}
\date{}
\begin{document}
\includepdf[pages=1-last]{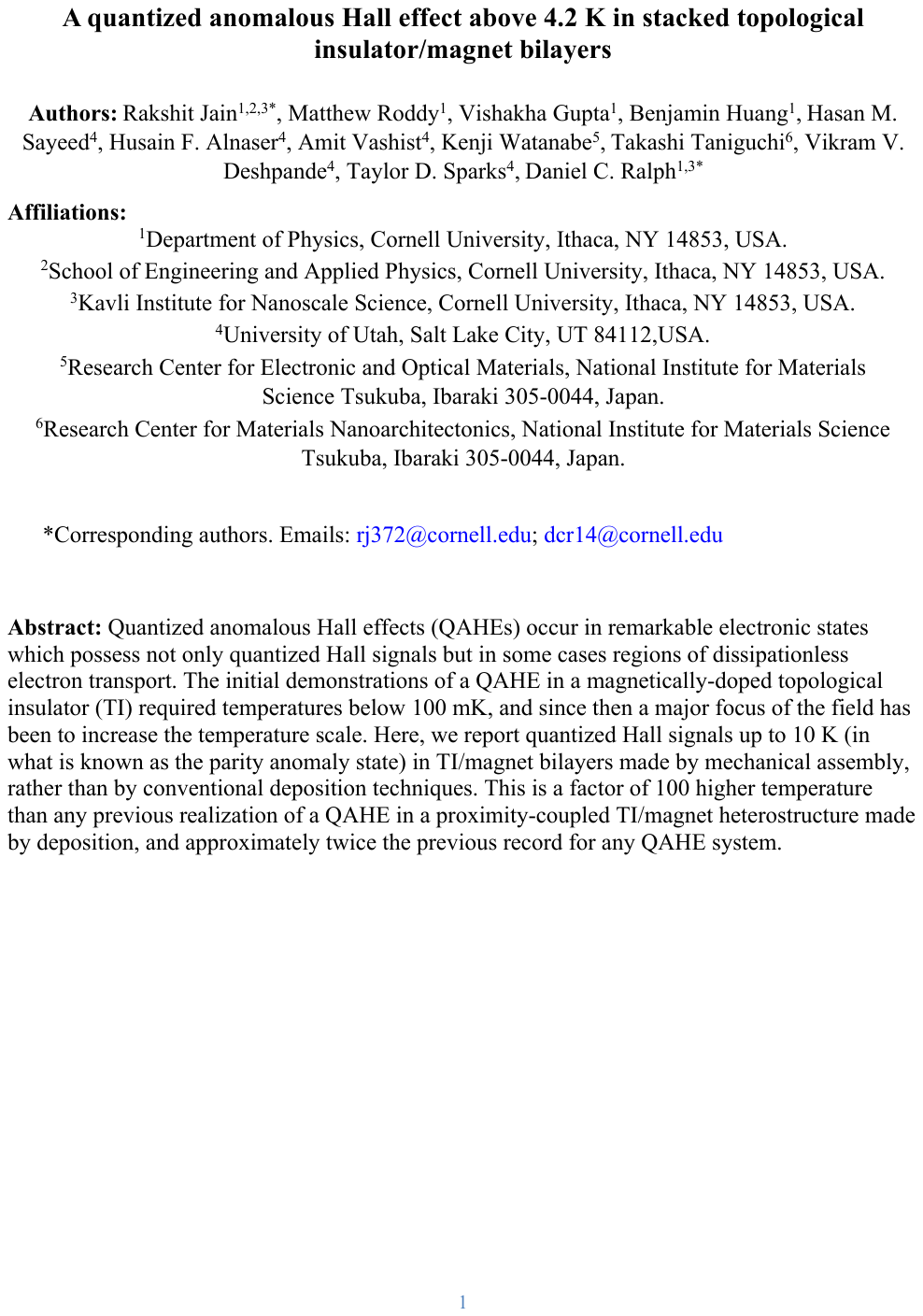}
\maketitle
\section{Materials and Methods}
\subsection{Steps for Device Fabrication}
\begin{figure}
    \centering
    \renewcommand{\thefigure}{S1}
    \includegraphics[width=\linewidth]{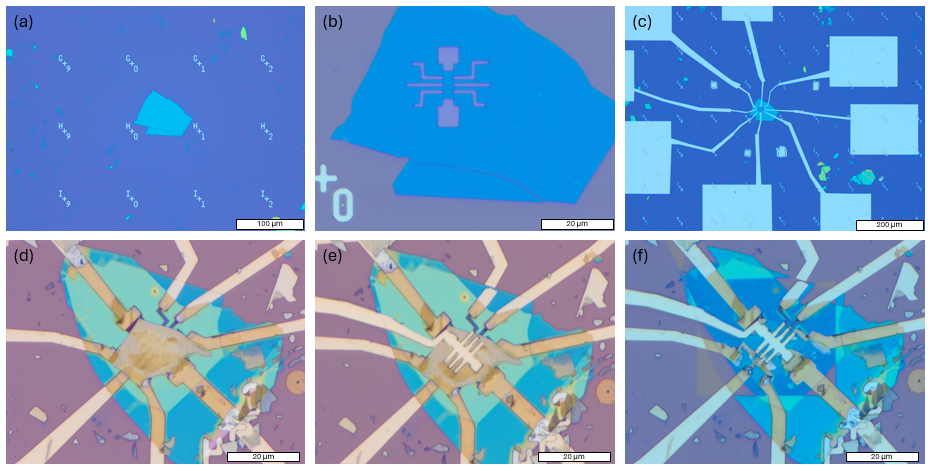}
    \caption{\textbf{Optical images of the fabrication process.} Images of the fabrication of electrodes for one device ((a)-(c)) and the subsequent creation of a top gate and etching into a Hall-bar geometry for a second device ((d)-(f)).  Scale bars are 100 $\mu$m in (a), 20 $\mu$m in (b, d, e, and f), and 200 $\mu$m in (c).
    }
    \label{fig:Fabrication}
\end{figure}

\begin{enumerate}
    \item Fabrication of Contact Electrodes
    \begin{enumerate}
        \item Exfoliate hexagonal Boron Nitride (hBN) onto pre-patterned alignment grids and identify suitable flakes using optical microscopy (\ref{fig:Fabrication}(a)).
        \item Design Hall-bar contacts with a channel length of 8\textmu m and a width of 3\textmu m to be contained on the hBN flake.
        \item For electron-beam lithography, spin-coat a bilayer of resists onto the wafer:
        \begin{enumerate}
            \item PMMA 495 A4, spun at 2000 RPM for 60 seconds then baked at 170$^\circ$ C for 2 minutes,
            \item PMMA 950 A4, spun at 2000 RPM for 60 seconds then baked at 170$^\circ$ C for 2 minutes.
        \end{enumerate}
        \item Perform electron beam lithography to pattern inner contact electrodes using a Zeiss Supra scanning electron microscope coupled with a Nabity nanometer pattern generating system with an acceleration voltage of 20 kV.
        \item Develop the resist using a chilled mixture of Isopropanol(IPA)/DI Water (3:1 ratio by volume).
        \item Deposit Pt(8 nm) for inner electrodes using an Angstrom electron-beam evaporator (\ref{fig:Fabrication}(b)).
        \item Lift-off excess metal in a bath of acetone followed by an IPA rinse.
        \item Repeat steps 1(c)-1(e) to pattern outer contact electrodes.
        \item Evaporate Ti(15 nm)/Pd(30 nm) for the outer electrodes and perform lift-off as previously described (\ref{fig:Fabrication}(c)).
        \item Clean the device channel and electrodes of residue by performing contact atomic force microscopy (AFM) over a large scan area ($\approx$ 40 \textmu m x 40 \textmu m).
    \end{enumerate}
    \item Material Exfoliation and Transfer
    \begin{enumerate}
        \item Exfoliate hBN onto Si/SiO$_2$ and identify appropriate flakes using optical microscopy.
        \item Exfoliate flakes of BiSbTeSe$_2$ (BSTS) and  Cr$_2$Ge$_2$Te$_6$ (CGT) using conventional scotch tape methods in an Argon filled glovebox with O$_2$ and H$_2$O concentrations below 0.1 ppm and identify flakes using optical microscopy.
        \item Pick up the flakes in top-to-bottom order (hBN $\rightarrow$ CGT $\rightarrow$ BSTS) using a dry-transfer polycarbonate/PDMS stamp and transfer to the contact electrodes (\ref{fig:Fabrication}(d)).
        \item Remove the polycarbonate by soaking in chloroform for $\approx$ 2 hours.
        \item Use AFM to measure the thickness of each layer in the device stack and ensure cleanliness of stack.
        \item Store the device in the inert Argon environment when not performing measurements or fabrication to minimize exposure to atmospheric conditions.
    \end{enumerate}
    \item Top Gate Deposition
    \begin{enumerate}
        \item Repeat steps 1(c) - 1(e) to pattern the top-gate in a design that covers the channel width and regions overlying contact electrodes.
        \item Deposit alternating layers of Ti(10 nm)/Pt(40 nm) 2 times in a sputtering system (\ref{fig:Fabrication}(e)).
        \item Lift-off excess metal using acetone, gently flushing with acetone to avoid damaging the gate, then rinsing in IPA.
    \end{enumerate}
    \item Etching into Hall-Bar Geometry
    \begin{enumerate}
        \item For lithography prior to etching, spin-coat a bilayer of resists onto the wafer:
        \begin{enumerate}
            \item PMMA 50 A8, spun at 2000 rpm for 60 seconds then baked at 170$^\circ$ C for 2 minutes,
            \item PMMA 950 A4, spun at 2000 rpm for 60 seconds then baked at 170$^\circ$ C for 2 minutes.
        \end{enumerate}
        \item Repeat steps 1(d) and 1(e) to pattern an etching region onto the device that removes any shorting regions of BSTS away from the channel and contact electrodes.
        \item Etch the hBN using a reactive ion etching tool: \begin{enumerate}
            \item Use a CHF$_3$/O$_2$ plasma at a chamber pressure of 50 mTorr and forward power of 50 W for 30 seconds,
            \item Examine the device under an optical microscope to determine if there is remaining hBN,
            \item Repeat the previous two steps 2-4 times depending on the thickness of the hBN until the hBN has been etched away entirely, revealing the exposed CGT and BSTS.
        \end{enumerate}
        \item Etch the CGT and BSTS using a wet etch bath composed of DI H$_2$O:H$_2$O$_2$:H$_3$PO$_4$ (Phosphoric Acid) in a 9:3:1 ratio by volume for approximately 1 minute then rinse in IPA.
        \item Examine the device under an optical microscope and if necessary, etch until the exposed BSTS and CGT is no longer visible (\ref{fig:Fabrication}(f)).
    \end{enumerate}
\end{enumerate}
\subsection{Measurements.}  The resistance measurements were carried out using a Keithley 6221 Source Meter to apply a 200 nA current bias at the frequency of 99.97 Hz, and measured using Signal Recovery Lockin Amplifier 7265 DSP. The top and bottom gate voltages were applied using Keithley 2400 source meters.  \\
\subsection{Capacitance Calculations.} The capacitance for the top gate is calculated using the capacitance of the top boron nitride flake and the CGT flake in series. 
\begin{equation*}
    C_{T} = \frac{C_\text{top-hBN} C_{CGT}}{(C_\text{top-hBN} + C_{CGT})}
\end{equation*}
The capacitance of the bottom gate is calculated using the capacitance of the SiO$_2$ and the bottom hBN flake in series.
\begin{equation*}
    C_{B} = \frac{C_{SiO_2} C_\text{bottom-hBN}}{(C_{SiO_2} +  C_\text{bottom-hBN})}
\end{equation*}
The capacitance of the individual layers is calculated using the $C = \frac{ k \epsilon_0}{d}$ where d is the thickness and k is the dielectric coefficient.
\subsection{Growth and characterization of the BiSbTeSe$_2$ crystals}
\subsubsection{Ampoule Preparation}
Bismuth (Bi), antimony (Sb), tellurium (Te), and selenium (Se) were sourced with a purity level of 5N grade from Sigma-Aldrich Co. To prepare the ampoule, quartz tubes (outer diameter of 1.2 cm, inner diameter of 0.8 cm) from Thermo Fisher Scientific Chemicals, Inc., were used. The quartz tubes were initially sealed at one end into a conical form, and the inner surface was coated with a carbon film through the pyrolysis of acetone. This coating acted as a protective barrier, preventing any potential reaction between the raw materials and the quartz during the heating process. The materials were weighed according to a 1:1:1:2 molar ratio of Bi:Sb:Te:Se, yielding a total mixture weight of 3 g. The prepared materials were placed inside the tube, which was subsequently flushed with argon gas multiple times to remove atmospheric air. Afterward, the tube was evacuated to create a vacuum environment, achieving a pressure below 10$^{-6}$ torr. Once the desired vacuum level was reached, the quartz tube was sealed, forming an ampoule with a final length of approximately 8–9 cm.

\subsubsection{Bridgman Growth}
The crystal growth was performed using a vertical Bridgman furnace equipped with three heating zones (\textit{48}). 
The ampoule, suspended vertically by a thin string connected to a motion controller, was positioned above the topmost heating zone. The temperatures of the furnace zones were set to 670°C (warm zone), 770°C (hot zone), and 500°C (cold zone), in descending order from top to bottom (\textit{38}). 
The temperature profile is illustrated in Fig.\ S2. The purpose of the top heating zone was to preheat the ampoule, preventing the condensation of Se vapor along the tube walls (\textit{49}). 
The ampoule was lowered into the furnace at a controlled rate of 0.6 cm/day, facilitating slow and controlled crystal growth along the vertical temperature gradient.

\subsubsection{Characterization of the Crystals}

Following crystal growth, the samples were carefully cleaved along their natural crystallographic planes. The exposed crystal surfaces were characterized using energy-dispersive X-ray spectroscopy (EDS) to determine the elemental composition. EDS measurements were carried out using an EDAX detector integrated with a FEI Quanta 600 scanning electron microscope (SEM), operated at an acceleration voltage of 15 kV. The technique provided high-resolution mapping of the distribution of Bi, Sb, Te, and Se elements across the sample.

\subsection{Temperature Dependence of Hall Conductivity}

We will consider the case that the Fermi energy is located in the middle of the exchange gap ($\Delta=$ full gap), so that at zero temperature the Hall conductivity has the saturated value for the parity anomaly state, $\sigma_{xy}(T=0) = e^2/2h$. As a function of increasing temperature, the magnitude of the Hall conductivity will decrease due to thermal activation of carriers from the valence band to the conduction band.  If we assume that the exchange gap and the form of the Berry curvature are temperature-independent, the change in the Hall conductivity as a function of temperature can be written as 

\begin{equation*}
    \Delta \sigma_{xy} = -2 \frac{e^2}{h} \int_{B.Z.} \frac{d^2 k}{(2\pi)^2} \Omega (k) f(E)
\end{equation*}
where $\Omega(k)$ is the Berry curvature as a function of wavevector $k$ and $f(E)$ the Fermi function as a function of energy.  The factor of two denotes the contribution from both unfilled and filled bands. Integration over angle in polar coordinates yields
\begin{equation*}
    \Delta \sigma_{xy} = -2 \frac{e^2}{\hbar} \int_0^{\infty} k dk \frac{2\pi}{(2\pi)^2} \Omega (k) f(E) = -2 \frac{e^2}{h} \int_0^{\infty} k dk \Omega (k) f(E).
\end{equation*}
The dispersion of the gapped surface state has the form 
\begin{equation*}
    E = \sqrt{(\hbar v_{f} k)^2 + \bigg(\frac{\Delta}{2}\bigg)^2 }
\end{equation*}
so that a change of variables from $k$ to the energy $E$ can be achieved via
\begin{equation*}
    dE = \frac{\hbar^2 v_f^2 k dK}{\sqrt{(\hbar v_{f} k)^2 + \big(\frac{\Delta}{2}\big)^2}} 
\end{equation*}
\begin{equation*}
    kdk = E dE \frac{1}{\hbar^2v_f^2}.
\end{equation*}
The Berry curvature depends on the exchange gap and carrier energy with the form 
\begin{equation*}
    \Omega = \frac{\hbar^2 v_{f}^2 \Delta}{4E^3}.
\end{equation*}
Thus the change Hall conductivity from the zero-temperature value can be expressed as an integral over energy in the form 
\begin{equation*}
   \Delta \sigma_{xy} = - \Delta \frac{e^2}{h} \int_{\frac{\Delta}{2}}^{\infty} dE \frac{f(E)}{E^2} = - \Delta \frac{e^2}{h} \int_{\frac{\Delta}{2}}^{\infty} dE \frac{1}{E^2 (1 + e^{\beta E})}
\end{equation*}
This can be simplified as 
\begin{equation}
    \Delta \sigma_{xy} = - \bigg(\frac{\Delta}{2}\bigg) \beta \frac{e^2}{h} \int_{(\frac{\Delta}{2})\beta}^{\infty} dx \frac{1}{x^2 (1 + e^x)}.
\end{equation}This final equation can be integrated numerically to obtain the predicted temperature dependence (again, this form assumes that the exchange gap is temperature-independent).  In Fig.\ S4 we have fitted this form to the measured data up to 15 K -- above 15 K the Hall conductivity begins to drop much more sharply than predicted by this simple form, indicating that the exchange gap is not temperature-independent. The fit indicates the temperature dependence of the Hall conductivity up to 15 K can be fitted reasonably with an exchange gap parameter of $\Delta= 4.5$ $\pm$ 0.5 meV . This value should be taken only as a lower bound on the zero-temperature exchange gap, because the gap may already be decreasing with increasing temperature below 15 K, and it is also possible that the Fermi energy was not truly centered in the middle of the exchange gap during the temperature-dependent measurements. This lower bound from the temperature-dependent measurements is therefore consistent with the exchange gap determined directly from the chemical-potential sensing measurement described in the main text, $\Delta(4.5$ K$)= 10 \pm 2$ meV.

\section{Data From an Additional Device}
Data from a second device with different dimensions are shown in Figure S3. The dimensions for this device are a width $w = 2$ $\mu$m and length $L = 3$ $\mu$m between the voltage probes used to measure the longitudinal resistance.

\newpage
\newpage
\begin{figure}
    \renewcommand{\thefigure}{S2}
    \centering
    \includegraphics[width=\linewidth]{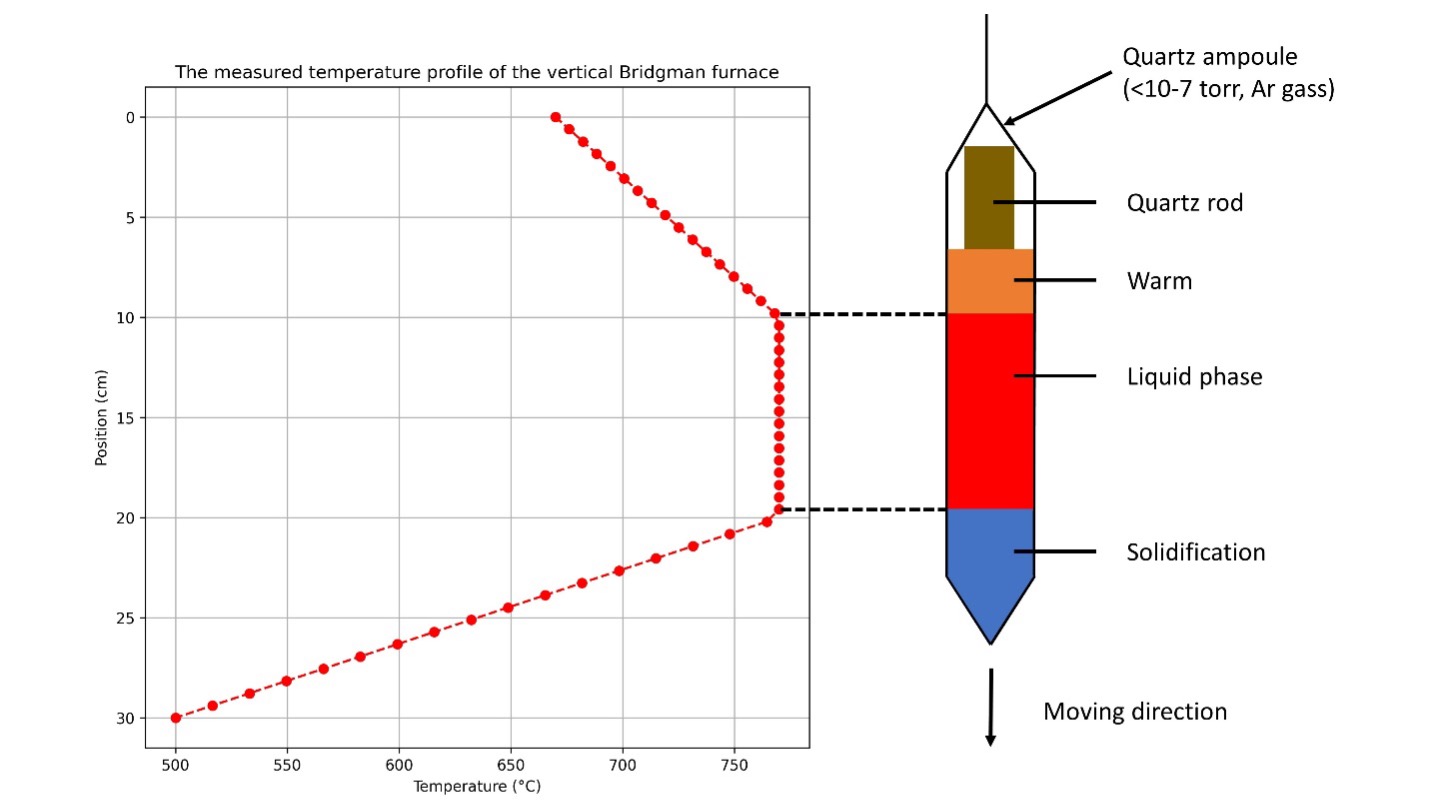}
    \caption{\textbf{The measured temperature profiles of the vertical Bridgman furnace}}
    \label{Fig S2}
\end{figure}
\FloatBarrier
\newpage
\begin{figure}
    \centering
    \renewcommand{\thefigure}{S3}

    \includegraphics[width=\linewidth]{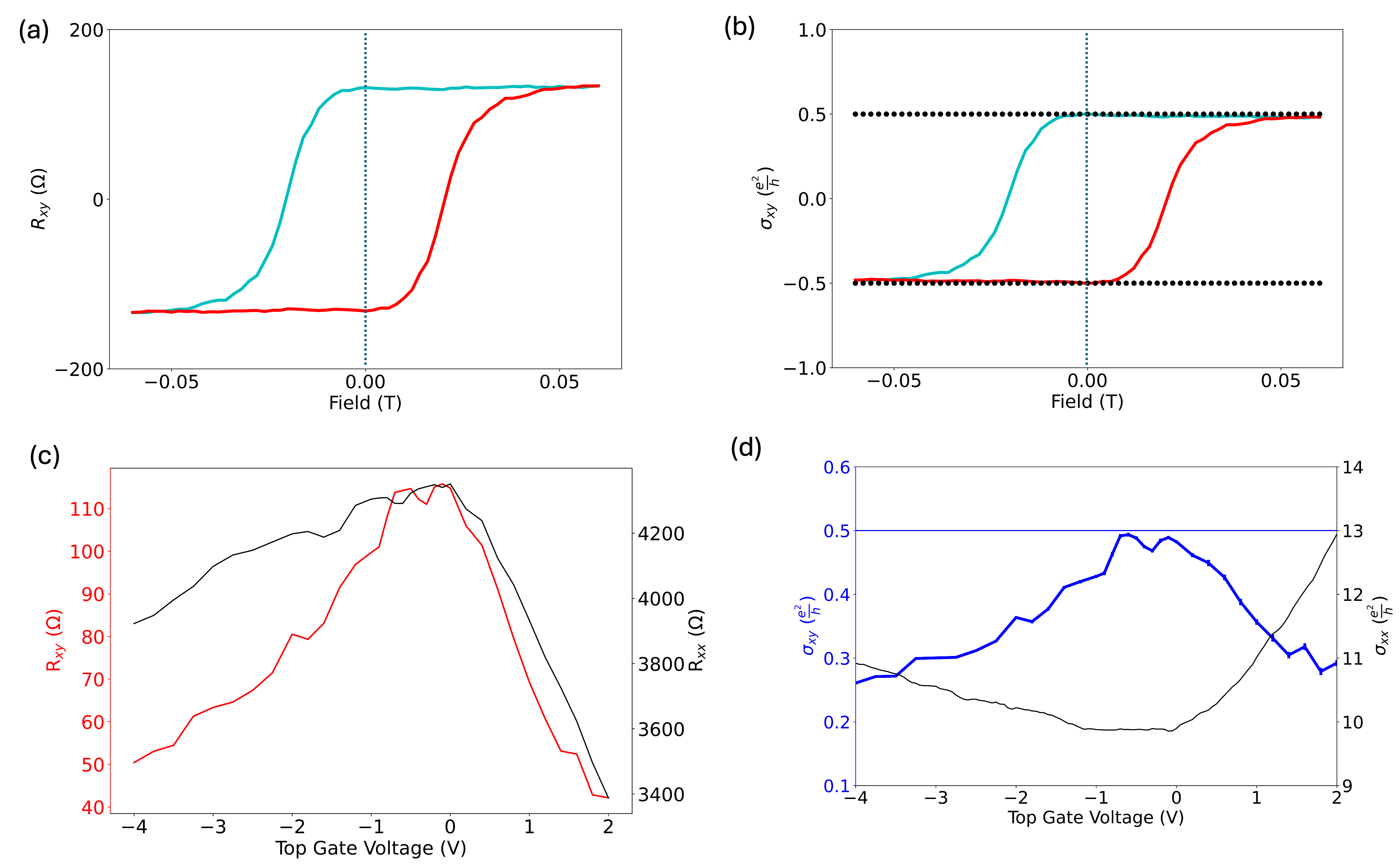}
    \caption{\textbf{Hall measurements at 4.5 K as a function of top gate voltage for a second device.} (a,b) Magnetic field sweep of the Hall resistance and Hall conductivity at a top gate voltage of $-0.1$ V, the value of $V_{TG}$ for which the chemical potential is near the middle of the exchange gap. (c) Hall and longitudinal resistances at 4.5 K and zero applied magnetic field as a function of  top gate voltage. (d) Hall and longitudinal conductivities at zero applied magnetic field as a function of top gate voltage. When the chemical potential lies inside the exchange gap, the Hall conductivity is quantized close to $e^2/2h$. The current excitation used for this measurement is 200 $\mu$A.
    } 
    \label{Fig S3}
\end{figure}
\FloatBarrier
\newpage

\begin{figure}
    \renewcommand{\thefigure}{S4}
    \centering
    \includegraphics[width=\linewidth]{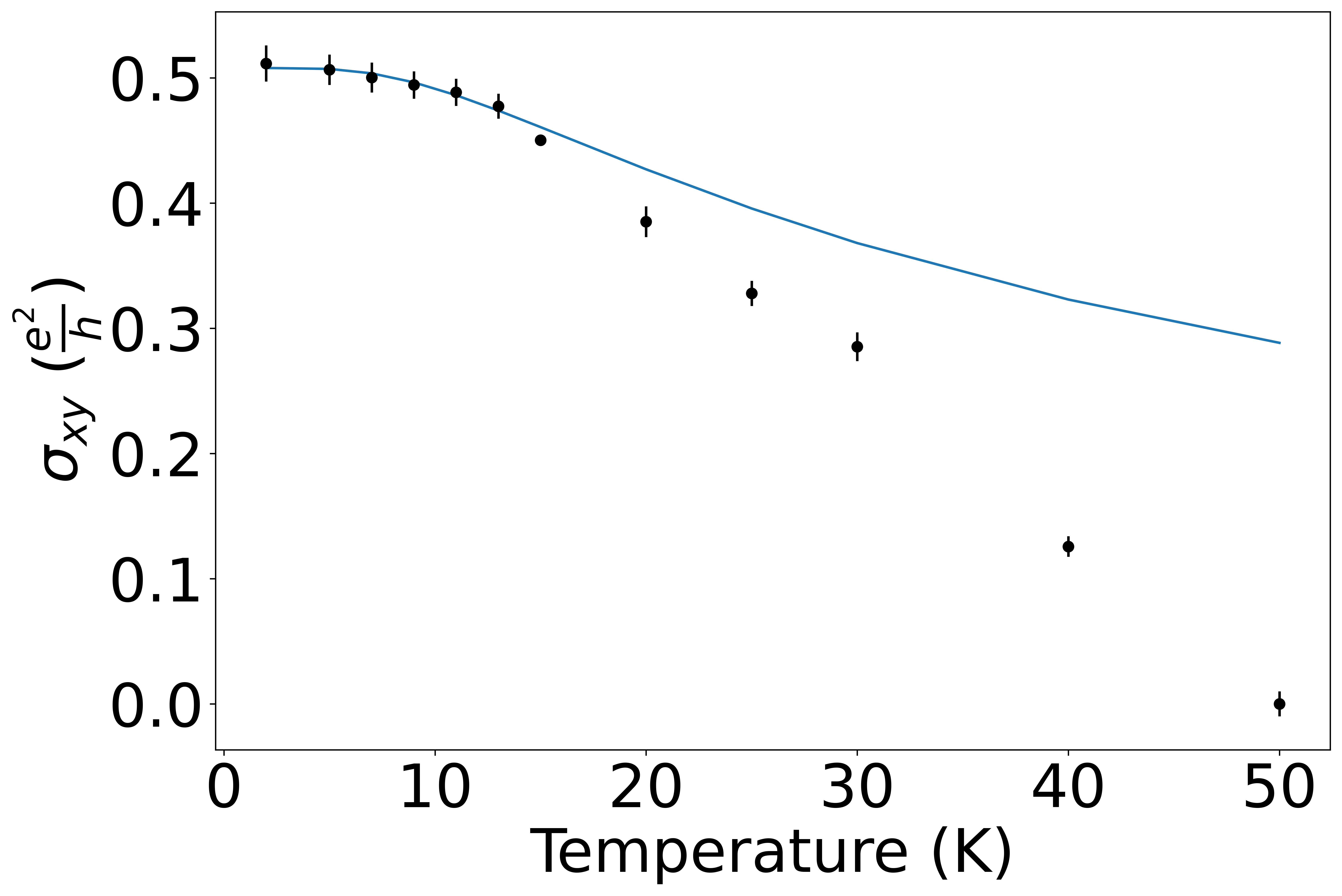}
    \caption{\textbf{Temperature dependence of the Hall conductivity} The circles are the data shown in the Fig.\ 4(b) in the main manuscript. The blue line is the fit of the data at temperatures $\leq 15$ K to Eq.\ (3) of the Supplementary Material. We estimate a lower bound for the low-temperature exchange gap of $\Delta  = 4.5 \pm 0.5$ meV from this fit.  (See Section 4 of the Supplementary Material for the details of the analysis.)}
    \label{Fig S3}
\end{figure}
\FloatBarrier
\newpage
\begin{figure}
    \renewcommand{\thefigure}{S5}
    \centering
    \includegraphics[width=\linewidth]{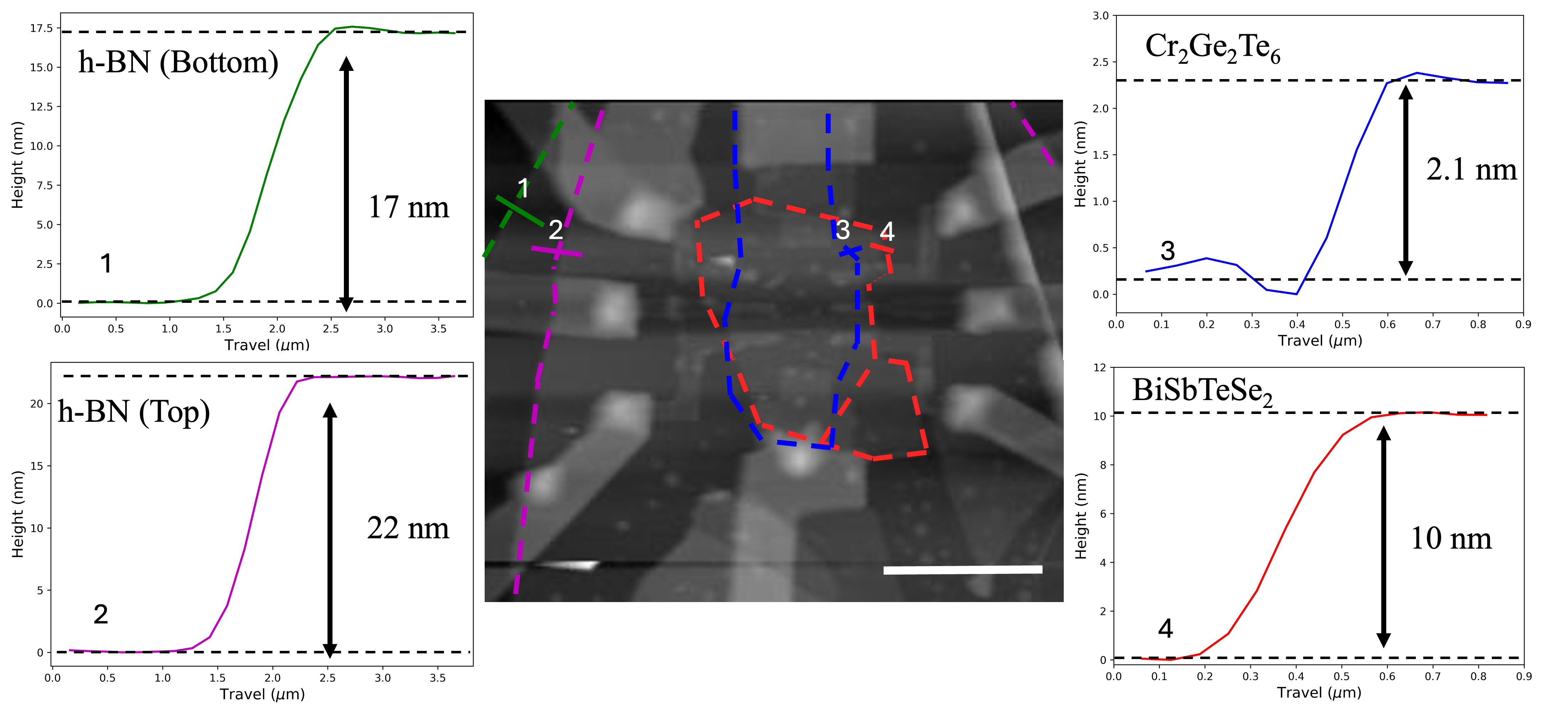}
    \caption{\textbf{Atomic force microscope image of the device analyzed in the main text.} The traces along 1, 2, 3, and 4 represent traces along the edge of bottom hBN, top hBN, Cr$_2$Ge$_2$Te$_6$, and BiSbTeSe$_2$ respectively. The scale bar for the image in the middle is 10 $\mu$m.}
    \label{Fig S4}
\end{figure}
\FloatBarrier

\begin{table}[h!]
\centering
\begin{tabular}{|c|c|c|}
     \hline
     Material & k (Dielectric constant) & d (thickness) \\
     \hline
     Top hBN~(\textit{50})& 3.4 & 22 nm\\
     \hline
     Cr$_2$Ge$_2$Te$_6$~(\textit{51})& 4 & 2.1 nm\\
     \hline
    BiSbTeSe$_2~(\textit{44})$ & 32 & 10 nm \\
     \hline
     Bottom hBN& 3.4 & 17 nm\\
     \hline
     SiO$_2$& 3.9 & 285 nm\\
     \hline
\end{tabular}
\caption{Dielectric coefficients for  materials used in the heterostructure.}
\end{table}

\end{document}